\documentclass{tlp}

\usepackage{iftex}
\usepackage{graphicx}
\usepackage{multirow}
\usepackage{graphicx}
\usepackage{flushend}
\usepackage[table]{xcolor}
\usepackage{amssymb}
\usepackage{amsmath}
\usepackage{hhline}
\usepackage{balance}
\usepackage{wrapfig}
\usepackage{framed}
\usepackage{lipsum}

\usepackage{textcomp}
\definecolor{PrologPredicate}{RGB}{0,0,200}
\definecolor{PrologVar}      {RGB}{145,032,039}
\definecolor{PrologComment}  {RGB}{169,082,044}
\definecolor{PrologOther}    {rgb}{0.2,0.2,0.2}
\definecolor{PrologString}   {RGB}{070,120,200}
\usepackage{relsize}
\usepackage{listings}

\newcommand{\code}{\lstinline[style=MyInline]}
\lstdefinestyle{MyInline}
{
    breaklines=true,
    breakatwhitespace=true,
    frame=lines,
  basicstyle = \ttfamily\color{PrologString},
}

\lstdefinestyle{MySCASP}
{
    xleftmargin=0.5cm,
    numberstyle=\tiny,
    numbers=left,
    stepnumber=1,
  mathescape = true,
  keywords = {},
  upquote = true,
  basicstyle = \ttfamily\color{PrologPredicate},
  basewidth = 0.47em,
  moredelim = {*[s][\color{PrologVar}]{(}{)}},
  moredelim = {*[s][\color{PrologOther}]{:-}{.}},
  commentstyle = \mdseries\color{PrologComment},
  morecomment=[l]\%,
}

\definecolor{linkcolor}{RGB}{52,59,144}
\usepackage[hidelinks, colorlinks=true, allcolors=linkcolor]{hyperref}
\usepackage{url}

\usepackage[textwidth=0.14\textwidth,textsize=scriptsize,backgroundcolor=none,
]{todonotes}
\ifpdf
  \usepackage{underscore}         
  \usepackage[T1]{fontenc}        
\else
  \usepackage{breakurl}           
\fi

\begin{document}

\jnlPage{X}{Y}
\jnlDoiYr{2024}
\doival{10.1017/xxxxx}

\title[Semantic Analysis of System Assurance Cases using Goal-directed ASP]{Automating Semantic Analysis of System\\ Assurance Cases using Goal-directed ASP}

\begin{authgrp}
  \author{%
    \sn{Murugesan}, \gn{Anitha}$^1$ \hspace*{.4cm}
    \sn{Wong}, \gn{Isaac}$^1$ \hspace*{.4cm}
    \sn{Arias}, \gn{Joaquín}$^2$ \hspace*{.4cm}\\
    \sn{Stroud}, \gn{Robert}$^3$ \hspace*{.4cm}
    \sn{Varadarajan}, \gn{Srivatsan}$^1$ \hspace*{.4cm}
    \sn{Salazar}, \gn{Elmer}$^4$ 
    \sn{Gupta}, \gn{Gopal}$^4$\hspace*{.4cm}
    \sn{Bloomfield}, \gn{Robin}$^{3,5}$ \hspace*{.4cm}
    \sn{Rushby}, \gn{John}$^6$
  }
  \vspace*{1em}
  \affiliation{%
    $^1$Honeywell Aerospace \hfill $^2$CETINIA, Universidad Rey Juan Carlos \hfill $^3$Adelard (part of NCC Group)\\
    $^4$University of Texas at Dallas \hfill~\hfill $^5$City, University of London\hfill~ \hfill \hfill\hfill $^6$SRI International \hfill~ 
    }
\end{authgrp}


\newcommand{\anitha}[1]{\textit{\textcolor{red}{Anitha:#1}}}

\maketitle

\begin{abstract}
    Assurance cases offer a structured way to present arguments and evidence for certification of systems where safety and security are critical. However, creating and evaluating these assurance cases can be complex and challenging, even for systems of moderate complexity. Therefore, there is a growing need to develop new automation methods for these tasks. While most existing assurance case tools focus on automating structural aspects, they lack the ability to fully assess the semantic coherence and correctness of the assurance arguments. 

In prior work, we introduced the Assurance 2.0 framework that prioritizes the reasoning process, evidence utilization, and explicit delineation of counter-claims (defeaters) and counter-evidence.  In this paper, we present our approach to enhancing Assurance 2.0 with semantic rule-based analysis capabilities using common-sense reasoning and answer set programming solvers, specifically s(CASP). By employing these analysis techniques, we examine the unique semantic aspects of assurance cases, such as logical consistency, adequacy, indefeasibility, etc. The application of these analyses provides both
system developers and evaluators with increased confidence about the assurance case.

\end{abstract}


\section{Introduction}
\label{sec:introduction}

Certification of systems in regulated industries, such as aerospace, nuclear, and healthcare, necessitates authorities to determine whether a system assuredly complies with domain standards such as security and safety, by evaluating the evidence of system assurance provided. However, as system complexity increases, the evidence presented using traditional, highly prescriptive, and process-driven approaches (such as DO-178C in aerospace) often becomes overwhelming and largely unstructured, which makes their compilation and review time-consuming and cumbersome. Hence, there has been a growing interest in recent years in developing \textit{Assurance Cases}, as an alternative means to present evidence and establish confidence in system compliance. 

The assurance case approach advocates for hierarchically structuring persuasive arguments, backed by a well-organized body of evidence, to effectively substantiate the top-level claim about the system, such as its compliance with standards. Many methodologies emphasize creating and presenting the assurance case using graphical formats. This systematic and visually engaging approach is increasingly recognized as a preferred choice for both organizations seeking certification and certifying agencies~(\cite{explicate78}). While assurance cases address challenges associated with systematically presenting large bodies of evidence, the descriptions used within their arguments and evidence are predominately natural language statements. Thus, interpreting the meaning of those statements to assess the arguments' coherence and evidence's relevance remains an intellectually demanding, manual task for assurance case authors and evaluators. 

Consequently, there is a significant drive towards researching and developing strategies to automate the creation and evaluation of assurance cases, aiming to reduce human effort and enhance confidence in the cases. While most current assurance tools and methodologies primarily concentrate on verifying the structural soundness of assurance cases, there is not adequate support to reason about the \textit{semantics} or the meaning conveyed by the natural language statements used within the case. 

To that end, DARPA's research program, Automated Rapid Certification of Software, was intended to explore and address the challenges associated with the generation and evaluation of assurance cases.  As part of this initiative, we have developed a comprehensive approach and tool suite named Consistent Logical Automated Reasoning for Integrated System Software Assurance (CLARISSA), which is founded on our assurance methodology called Assurance 2.0 (described in \cite{bloomfield2020assurance}).
While the complete details of CLARISSA are elaborated in \cite{finalCLARISSAReport}, \cite{murugesan2023semantic}, \cite{defeatersreport}, and \cite{bloomfield2024assessing}, we provide only brief and informal introductions here.  

In this paper, we present a subset within our CLARISSA approach that automates the analysis of the semantic aspects of Assurance 2.0, by leveraging the power of common sense reasoning and answer set programming (ASP) \cite{brewka2011answer,gelfond2014knowledge} solvers, namely s(CASP)~\cite{scasp-iclp2018}. s(CASP) is a novel goal-oriented, non-monotonic reasoner capable of efficiently handling logical reasoning (see Section~\ref{sec:scasp-non-monotonic}) tasks essential for semantic analysis, which is central to our research. The goal-directed execution of s(CASP) automates commonsense reasoning based on general rules and exceptions (through default negation). That is, it enables inferences to be drawn from a set of logical rules that formalize the assurance case. Additionally, s(CASP) performs \textit{deductive and abductive reasoning}, which is essential for proving whether a top-level claim can be deduced based on the arguments, evidence, and/or assumptions in assurance cases, and make it possible to define invariants (as \textit{global constraints}) that allows the analysis of scenarios that violate these invariants.

In a nutshell, we leverage the general pattern of each assurance statement -- that intuitively is nothing but assertions of \textit{properties} of \textit{objects} in a certain \textit{environment} -- to perform semantic analysis. We first capture crucial lexical components -- namely objects, properties, and environment -- within the assurance case statement using minimal formalized semantics. Subsequently, we automatically translate the formalized elements of the entire assurance case into corresponding logical rules under Answer Set Programming (ASP) semantics. Using various semantic properties of interest expressed as rules and queries, we use s(CASP) to provide the proof that these properties (represented as predicates) hold in the assurance case.

To the best of our knowledge, this methodology is not only the first attempt to tackle the automated semantic analysis of system assurance cases, it also establishes a new paradigm for explainable, knowledge-assisted assurance reasoning. This is thanks to the top-down solving strategy utilized by s(CASP), which produces concise and human-understandable justifications (\cite{arias-justification}). These justifications are essential to precisely identify the reasons for the success (positive queries) or failure (negative queries) of the assurance so that we can resolve the issues raised. 
Furthermore, s(CASP) supports various forms of negation, each with unique applications in assurance reasoning. As we mentioned before, we use \textit{default negation} to derive detailed justifications for why a given claim cannot be proven, while \textit{classical} or {strong negation} can be used to impose specific restrictions (see the book by \cite{gelfond2014knowledge} for details).

We have evaluated our approach on multiple industrial strength case studies, particularly in the fields of Avionics and Nuclear Reactors (refer \cite{finalCLARISSAReport}). In this paper, we use one of the case studies, the ArduCopter system, an open-source platform (see Section~\ref{sec:case-study:-arduc}), to illustrate the approach and highlight the contributions:
\begin{itemize}

\item Methodically identify and capture the vocabulary within each assurance case statement that contributes to the lexical significance of the assurance case. The main steps of this methodology are outlined in Section~\ref{sec:semant-analys-appr}. 

\item Systematically translate the assurance case into an answer set program that is amenable for execution by the s(CASP) engine. To facilitate this, we have enhanced the ASCE tool with a plugin that captures the vocabulary and automatically transforms the use case into logical notation. The new plugin is described in Section~\ref{sec:form-assur-stat},  and the transformation mapping is explained in Section~\ref{sec:transf-logic-notat}.
    
\item Identify and formalize properties that are unique and essential for assessing the semantic rigor of assurance cases. These properties are elaborated in Section~\ref{sec:semant-prop-defin}. 

\item Leverage the capabilities of s(CASP) tool to analyze semantic properties within the assurance case, as described in  Section~\ref{sec:semant-analys-using}. In particular, our use of \textit{Negation as Failure} involves deriving negative conclusions from the absence of positive information, to automatically identify defeaters. Additionally, s(CASP) possesses the ability to perform \textit{non-monotonic reasoning} that allows revision of conclusions in light of new information. This feature is particularly valuable for incrementally assessing and improving the strength of assurance cases during the authoring process.

\end{itemize}

\section{Motivation}

\begin{figure}
\begin{center}
    \includegraphics[width=0.9\columnwidth]{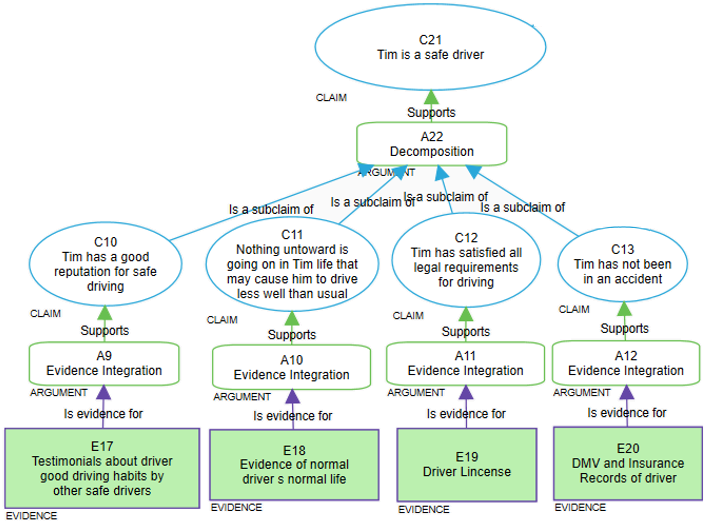}
    \caption{Safe Driver - A Motivating Example}
    \vspace{-0.15in}
    \label{fig:safedriver}
    \end{center}
\end{figure}

Consider the scenario, used in ~\cite{holloway2015understanding}, where an assurance case is constructed with the intent to convince Jon's father that Tim (a college student known to the family) is a safe driver to take Jon (a teenager not yet of driving age) to a football game in Tim's car. The assurance case, as shown in Figure~\ref{fig:safedriver} (uses Assurance 2.0 format that is elaborated in the next section), has a top-level claim \textit{``Tim is a safe driver''} (top-most blue ellipse), that is supported by several fine-grained sub-claims (lower level blue ellipses) and corroborating evidence (purple rectangles at the leaf-level) that establishes Tim’s capability to drive safely. Jon's father must evaluate the assurance case and determine if he is convinced. If not, he must explain. 

Evaluating assurance cases is a complex process that involves several critical steps. It requires analyzing the semantics or meanings of the assurance statements to verify that sub-claims are articulated accurately and consistently, that they collectively support the overarching claim, and that the provided evidence is relevant and sufficiently substantiates the claims. Any semantic discrepancies or gaps discovered during the evaluation undermine confidence in the overall claim. For example, if testimonials (E17 in the Figure) indicate Tim's good driving but also mention a minor road incident not formally recorded by the DMV, it contradicts the sub-claim (C13 ellipse in the Figure) asserting Tim has not been involved in any accidents. Similarly, if Tim's driver's license was issued in a state different from where the football game took place, this discrepancy indicates that the context or environment in which the evidence is presented does not adequately support the top-level claim. Furthermore, the absence of claims and evidence concerning the condition of Tim's car in the assurance case, which is crucial for fully convincing Jon's dad, further diminishes confidence. As exemplified using a few semantic gaps, the evaluation process involves interpreting the meaning of statements both individually and in conjunction with other statements from various perspectives.

Currently, no tools or techniques are available to support evaluators of assurance cases, such as Jon's dad, necessitating manual and labor-intensive evaluation processes that are prone to errors. The development of automation to aid in the semantic analysis of assurance cases—by identifying gaps such as inconsistencies among statements, inadequacies in claims, the validity of evidence, and deficiencies in the assurance strategy—would be highly advantageous. Evaluating the semantic aspects of assurance cases for even moderately complex real-world systems is both time-consuming and intellectually demanding. Automated semantic analysis tools would significantly alleviate the cognitive load on evaluators and could also assist authors in enhancing the reliability of their assurance cases. The remainder of this paper outlines our approach to providing automated support for the semantic analysis of assurance cases developed based on Assurance 2.0 principles.

\section{Background}
\label{sec:background}


In this section, we present a brief introduction to: (i) Assurance 2.0, the framework/methodology used to create assurance cases of engineering systems, 
and (ii) s(CASP), the reasoning engine upon which the semantic analysis of our proposal relies.

\subsection{The Assurance 2.0 framework}
\label{sec:fram-assur-2.0}

The Assurance 2.0, defined by \cite{bloomfield2020assurance}, is a modern framework that supports reasoning and communication about the behavior and trustworthiness of engineered systems and their certification. It provides a framework that separates out the deductive and inductive reasoning combined with the use of a practical indefeasibility criterion for justified belief. This frames the notion of defeaters, both \textit{undercutting} and \textit{rebutting}, and motivates construction of arguments that are predominately deductive, an approach known as ``Natural Language Deductivism”.  Details on \textit{defeaters} and \textit{eliminative argumentation} are provided by \cite{defeatersreport}. We also advocate the use of \textit{Confirmation Measures} to evaluate the strength of evidence and arguments. We  reduce confirmation bias through active search for defeaters and a methodology for doing so by means of counterclaims and counter-cases. We argue in \cite{bloomfield2024assessing} that confidence cannot be reduced to a single attribute or measurement. Instead, we draw on three different perspectives: positive, negative, and residual doubts. Our work also provides details of the approach to logical evaluation and soundness.

This framework adopts a Claims-Arguments-Evidence (CAE) approach with an increased focus on the evidence, doubts/objections, and reasoning and overall semantics of a case. The building blocks of a typical Assurance 2.0 case are claims (and sub-claims) that assert the properties of objects, such as ‘The train is safe’; evidences which are artifacts establishing trustworthy facts directly related to a claim; arguments that serve as bridging rules connecting what is known or assumed (sub-claims, evidence) to the claim under investigation; side claims which offer additional justification or assumptions to support the argument; defeaters that capture doubts and objections that challenge the validity of claims, arguments, or evidence; and Theories defined as reusable templates that can be instantiated in concrete assurance cases as sub-cases. Assurance 2.0 cases are authored using the Assurance and Safety Case Environment (ASCE) tool developed by Adelard LLP (2024). Figure~\ref{fig:cae} in Section~\ref{sec:case-study:-arduc} shows an exemplar of an assurance case authored in ASCE tool using Assurance 2.0 principles.

\subsection{s(CASP), a non-monotonic reasoner}
\label{sec:scasp-non-monotonic}


To conduct the semantic analysis, we leverage the advances in the field  of logic programming. In particular, we use Answer set programming (ASP), a paradigm rooted in logic programming (see paper by \cite{brewka2011answer} for details), which integrates reasoning methods that automates commonsense reasoning capturing and managing incomplete information, cyclical reasoning, and constraints.

Among the different ASP solvers, we chose s(CASP), a goal-directed
ASP system that executes answer-set programs in a top-down manner.
%
 The goal directed execution of s(CASP) is particularly well-suited for reasoning about assurance cases, because it generates partial stable models including only the relevant information needed to support (or decline) a given claim.
 Assurance cases are commonly structured in a way that makes sense to human interpretation.
 \textit{Deductive and abductive reasoning}, supported by s(CASP), is essential for proving whether a top-level claim of an assurance case can be deduced based on the arguments, evidence, and assumptions.
Both forms of negation are supported by s(CASP), and they have diverse applications in assurance reasoning.
 For instance, we  use \textit{default negation} to derive detailed justifications for why a given claim cannot be proven.
 Additionally, we are exploring the use of \textit{abducibles}, which involves \textit{even loops over negation} and makes it possible to derive negative conclusions from the absence of positive information, and  automatically identify defeaters.
 s(CASP) supports reasoning about \textit{global constraint}s and \texttt{classical negation}, so it is able to analyze scenarios that violate these constraints.
 Furthermore, the top-down solving strategy employed by s(CASP) generates concise, human-understandable \textit{justifications} (see work by \cite{arias-justification} for details).
 These justifications play a crucial role in precisely identifying the reasons for assurance failure and resolving concerns.
 Lastly, s(CASP) possesses the ability to perform \textit{non-monotonic reasoning}, allowing for the revision of conclusions in light of new information (due to the presence of negation).
 This feature is particularly valuable for incrementally assessing and improving the strength of assurance cases during the authoring process.

\section{Illustrative Case Study: The ArduCopter system}
\label{sec:case-study:-arduc}

To demonstrate the concepts of our approach and tools, we will utilize the open-source ArduCopter system as a case study in the remainder of this paper.  This system was previously utilized to assess the CLARISSA methodology. Additional evaluations conducted on industrial-strength case studies from the avionics and nuclear sectors are detailed in \cite{finalCLARISSAReport}.

\begin{figure}[!h]
\begin{center}
    \includegraphics[width=1\columnwidth]{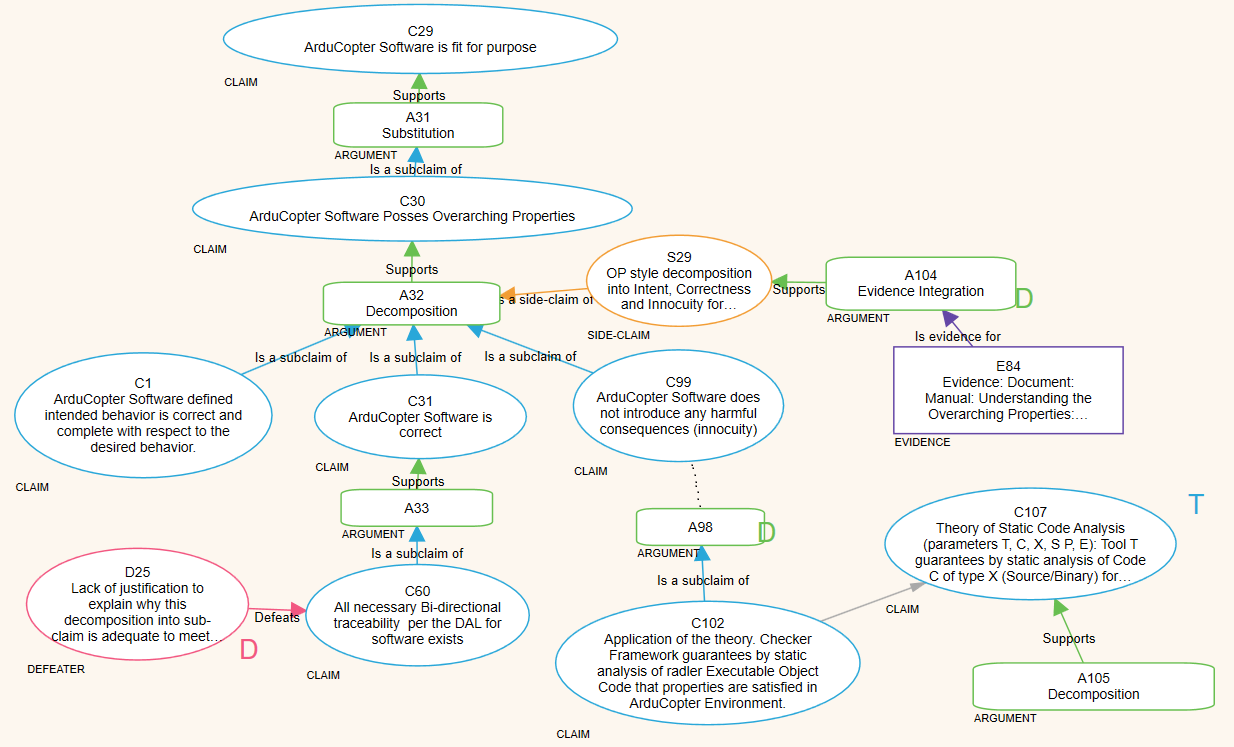}
    \caption{Arducopter Case Fragment in Assurance 2.0}
    \label{fig:cae}
    \end{center}
    \vspace{-0.3in}
  \end{figure}
  
 The ArduCopter, derived from the open-source ArduPilot autopilot platform by \cite{ardupilot}, is an unmanned aerial vehicle designed to oversee a diverse range of avionic vehicles, enabling them to perform various autonomous tasks. Our focus with ArduCopter is to construct an assurance case using the Assurance 2.0 methodology, establishing justified confidence in its ability to conduct autonomous surveillance missions while adhering to safety and security standards. The complexity of the available details, including the concept of operations, system architecture, and other development and verification artifacts for the ArduCopter system, was sufficient to construct a realistically sized assurance case with an appropriate level of complexity. This allowed us to explore the associated challenges and assess the effectiveness of our approach. We will use excerpts from our evaluation of this system to illustrate the principles of our approach.

Figure~\ref{fig:cae} depicts a snippet of the ArduCopter assurance case constructed using the ASCE tool. The overall structure of the assurance case aims to demonstrate that the ArduCopter exhibits the three fundamental overarching properties—intent, correctness, and innocuity —that are indispensable for ensuring the safety and security of the system.  The top-level \textit{claim} of ArduCopter (the uppermost blue ellipse) is broken down into detailed \textit{sub-claims} (blue ellipses) through the use of \textit{arguments} (green rounded rectangles), which are substantiated by \textit{evidence} (purple rectangles) at the leaf level. The rationale behind these refinements is documented in \textit{side-claims} (yellow ellipses). Moreover, any doubts, concerns, or counter-claims regarding any aspect of the case are captured as defeaters (red ellipses), described by \cite{defeatersreport}. Several theories (blue ellipses adorned with a ``T'' symbol) were employed in formulating the ArduCopter assurance case, such as the theory of static analysis (ID C107) and its application in a sub-claim (ID C102). The complete details of this case are available in \cite{finalCLARISSAReport}.

 \vspace{-0.15in}
\section{Semantic Analysis Approach}
\label{sec:semant-analys-appr}

To a large extent, assurance cases heavily rely on unstructured, free-form natural language despite their structured graphical representation, which is not naturally conducive to automation. On the contrary, complete formal notations allow for automated analysis, but they present challenges in authoring and reviewing without a steep learning curve and also have expressibility limitations. Seeking a middle ground, we defined our approach that facilitates semantic analysis by capturing essential lexical components of assurance statements in an intuitive and ``minimally'' formal manner.

\begin{figure}[!h]
\begin{center}
    \includegraphics[width=1\columnwidth]{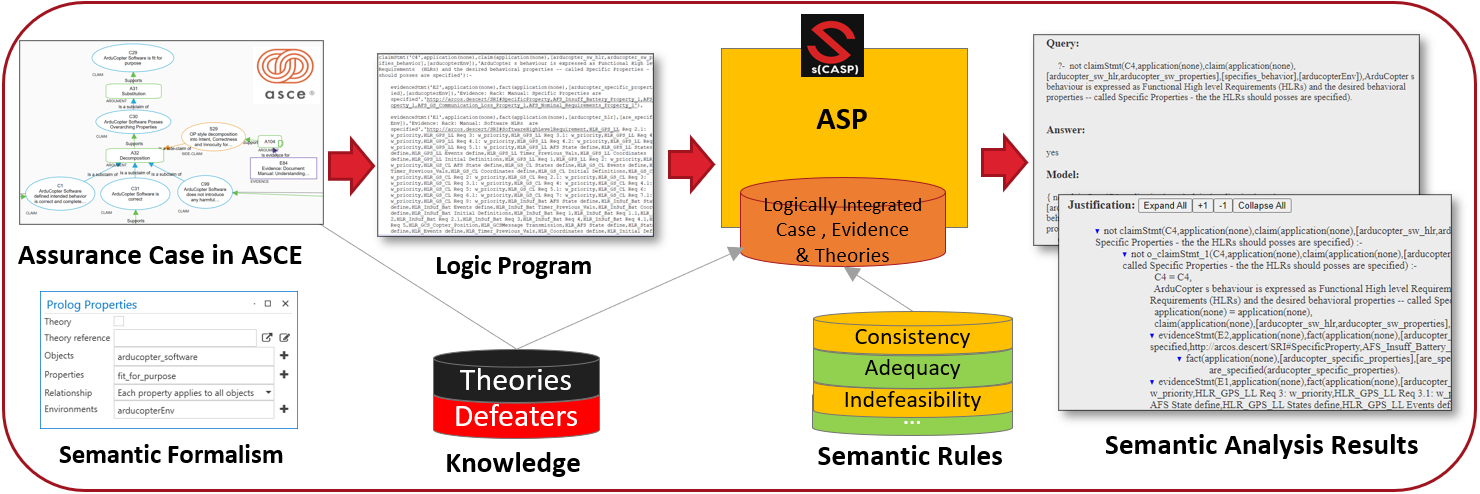}
    \caption{Semantic Analysis Approach}
    \vspace{-0.15in}
    \label{fig:approach}
    \end{center}
\end{figure}

Our approach to semantic analysis, shown in Figure~\ref{fig:approach}, has the following steps:  
\begin{enumerate}
    \item \textbf{Formalism of Assurance Statements}:  The crucial language components used to articulate statements within Assurance 2.0 cases are categorically grounded and formally captured within the ASCE tool interface.
    \item \textbf{Transformation into ASP}:  The formalized language along with the assurance case structure is automatically transformed and exported into equivalent logical predicates in ASP.
    \item \textbf{Modeling Semantic Properties}:  Various properties that ensure semantic rigor of Assurance 2.0 cases are modeled as rules.
    \item \textbf{Semantic Reasoning using s(CASP)}: The predicate form of the assurance case is systemically analyzed for the various properties using s(CASP) engine and results are reported in a user-friendly manner.  
\end{enumerate}
While Figure~\ref{fig:approach} is aimed to provide a comprehensive high-level overview of our approach, in the following subsections we elaborate on each step and offer details of the text presented in smaller font within the figure.


\subsection{Formalism of Assurance Statements}
\label{sec:form-assur-stat}

Assurance cases consist of blocks or ``nodes'' that delineate the
properties or relationships relevant to objects within a specific
environment. The ``Properties,'' ``Objects,'' and ``Environment'' for
each node are identified and defined, enabling us to formally
articulate statements like ``Object O satisfies property P in
environment E.'' For instance, the assertion ``ArduCopter Software is
fit for purpose in its Intended Environment'' can be broken down into
(i) ``Object = ArduCopter Software'', (ii) ``Property = fit for
purpose'', and (iii) ``Environment = Intended Environment
(arducopterEnv).''

\begin{figure}[ht]
\begin{center}
    \includegraphics[width=1\columnwidth]{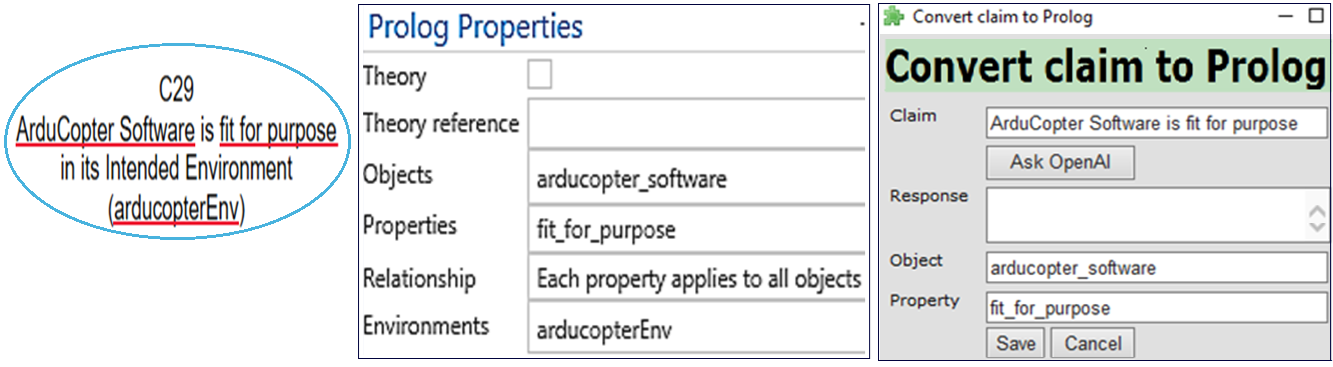}
    \caption{Objects-Properties-Environments Formalism and LLM Support}
    \label{fig:ope}
    \end{center}
\end{figure}

Currently, the ASCE assurance case authoring tool allows users to explicitly record objects, properties, and environments as formal semantics, along with their relationships as depicted in Figure~\ref{fig:ope}. Besides offering a field for natural language descriptions, which aids in the manual inspection of the assurance case, it also allows manual entry of these formal semantics by users. These Objects, Properties, and Environments will be crucial for the next step of transforming the assurance case into a logic program. Because it can be tedious and time consuming to enter these semantics for every node, we have enhanced the ASCE tool with open-source LLMs to parse the natural language description and automatically extract these semantics. Although the current LLM support is limited, we are investigating methods to improve their use.


\subsection{Transformation into ASP}
\label{sec:transf-logic-notat}

Based on the formalism captured for each node of the assurance case, we have augmented the ASCE tool with a plugin to automatically transform the entire assurance case into equivalent logical predicates in ASP. The choice to transform into ASP is based on the fact that each concept in Assurance 2.0 and the intended analysis can be readily mapped to corresponding concepts in first order logic. The mapping between these concepts is illustrated in Table~\ref{tab:ass2scasp}.

\begin{table}[ht]
\begin{center}
\rowcolors{2}{gray!10}{gray!0.5}
\begin{tabular}{|p{2.3cm}|p{10.5cm}|}
\rowcolor{gray!40}
\hline \textbf{Assurance 2.0 Concept} & \textbf{Mapping into first order logic}\\ \hline 
{Top-level Claim} & In the assurance case, the top-level claim aligns with the concept of \textit{queries}, which can be subjected to logical entailment analysis.\\\hline 
{Sub Claims and Side Claims} & Sub-claims and side claims are analogous to \textbf{predicates}  serving the purpose of capturing the relationships between objects.\\\hline 
{Argument}  & Arguments that establish relationships among sets of objects-properties-environment can be equated to \textbf{rules} where logical implications are utilized to describe these relationships. \\  \hline 
  {Evidence} &  Evidence artifacts that establish the truth about the system are \textbf{facts}. \\\hline 
{Defeater} &  Defeaters are counter-claims to provide support for not believing that claim, which is the same as \textbf{negated goals} in formal logic.\\ \hline 
{Binding} & Theory definitions require the specification of variables for objects and the environment, that will be instantiated with terms when applied to a specific assurance case. This process aligns with the concept of \textbf{substitution}, which involves dynamically mapping variables to terms. \\ \hline 
{Reasoning} & Assurance case reasoning involves demonstrating that a top-level claim is entailed based on its arguments and evidence, akin to \textbf{proofs} in logic, where the validity or truth of a claim is established through logical reasoning. \\ \hline 
\end{tabular}
\caption{Concept Mapping between Assurance 2.0 and first order logic.}
\label{tab:ass2scasp}
\end{center}
\end{table}

\begin{table}
    \begin{tabular}{|p{1.6cm}|p{11.4cm}|}
      \rowcolor{gray!40}
      \hline
      \textbf{Node}
      & \textbf{Translation into ASP} \\
      \hline 
      {Claim}
      & \vspace{-1em}
\begin{lstlisting}[style=MySCASP, numbers=none, xleftmargin=0cm]
claimStmt(Claim_ID, Application, ClaimPredicate, Description) :-
    [claimStmt | evidenceStmt | side_ClaimStmt], $\dots$, [-Defeater],
    ClaimPredicate, 
    theory(Theory_ID, Application, Claim).
ClaimPredicate :- 
    PropertyList.
theory('Theory_ID', Application, ClaimPredicate) :-
    [claimStmt | evidenceStmt | side_ClaimStmt], $\dots$, 
    ClaimPredicate.   
\end{lstlisting}
      \\
      \hline
      {Evidence}
      & \vspace{-1em}
\begin{lstlisting}[style=MySCASP, numbers=none, xleftmargin=0cm]
evidenceStmt(Evidence_ID, Application, ClaimPredicate, Artefact, URI) :-
    [-Defeater],  ClaimPredicate.
ClaimPredicate :- 
    PropertyList. 
\end{lstlisting}
      \\
      \hline
      {Side Claim}
      & \vspace{-1em}
\begin{lstlisting}[style=MySCASP, numbers=none, xleftmargin=0cm]
side_ClaimStmt(Side_Claim_ID, Application, ClaimPredicate, 
                                                       Justification) :-
    [claimStmt | evidenceStmt | side_ClaimStmt], $\dots$, [-Defeater], 
    ClaimPredicate.
ClaimPredicate :- 
    PropertyList. 
\end{lstlisting}
      \\
      \hline   
      {Defeater}
      & \vspace{-1em}
\begin{lstlisting}[style=MySCASP, numbers=none, xleftmargin=0cm]
defeater(Defeater_ID, Application, defeats(Node_ID), 
                                         ClaimPredicate, Description) :-
    [claimStmt | evidenceStmt | side_ClaimStmt], $\dots$, [-Defeater],
    [theory('Theory_ID', Application, Claim)].
ClaimPredicate :- 
    PropertyList.
PropertyList.
\end{lstlisting}
      \\
      \hline
    \end{tabular}
    \caption{Assurance 2.0 Node Mapping into ASP.}
    \label{tab:ASCEtoProlog}

\end{table}

\begin{figure}[!h]
\begin{center}
    \includegraphics[width=0.8\columnwidth]{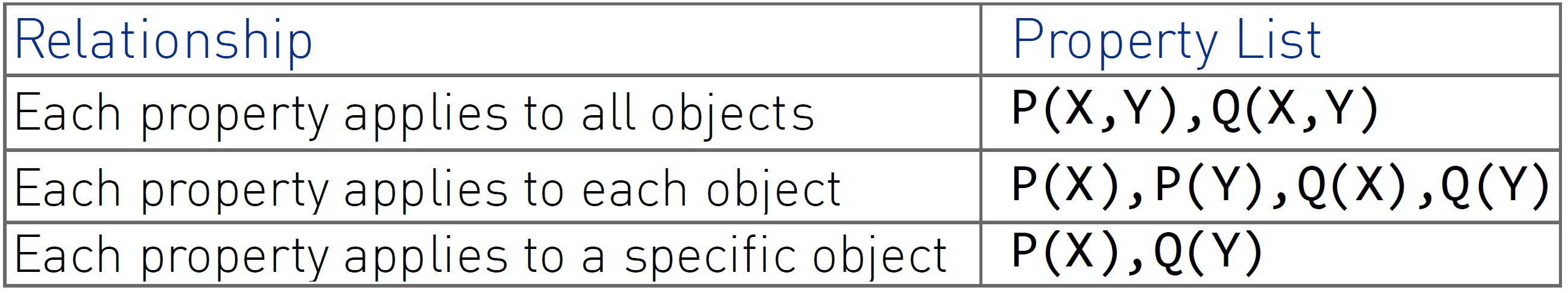}    
    \caption{Object-Property Relationships}
    \label{fig:Relationship}
\end{center}
\end{figure}

Table \ref{tab:ASCEtoProlog} shows the rule to transform each node type to predicate form. The term \code{ClaimPredicate} refers to a ASP predicate represented as the \code{claim([O],[P],[E])}, where \code{[O]}, \code{[P]}, and \code{[E]} represent comma-separated lists of ``Objects'', ``Properties'', and ``Environments'' associated with each assurance node. As we formalize the properties, the relationships specified between lists of objects and properties are preserved. This preservation ensures precision in the analysis. ASCE tool allows 3 different types of relationship specification, in addition to the `Off' to indicate no relationship. Consequently, the \code{PropertyList} is derived from the \code{ClaimPredicate} based on the generic object-property relationship definition, as described in Figure~\ref{fig:Relationship}. If the Relationship property is Off, which is its default value, the PropertyList is empty and the ClaimPredicate is asserted as an ASP fact, without expanding the property list.  For example, if properties such as {\it consistent} and {\it verifiable} are specified for both high and low-level requirements specification artifacts, the property list is formulated as \code{consistent(high-level-requirements, low-level-requirements)} and \code{verifiable(high-level-requirements, low-level-requirements)}. On the other hand, if the properties such as {\it traceable to design} and {\it traceable to test cases} are asserted for specific objects of high and low-level specification artifacts, respectively, the property list will expand to \code{traceable-to-design(high-level-specification)} and \code{traceable-to-test-cases(low-level-specification)}. Hence, depending on the specified relationship between the properties and objects, the \code{ClaimPredicate} is  expanded in ASP syntax and used for the analysis. 
Furthermore, to maintain the structure of the case during export and ensure traceability, certain metadata (such as node identifiers, descriptions, etc.) is also included within the exported predicates. Figure~\ref{fig:PrologExport} shows an
example of the exported predicates of Arducopter assurance case.

\begin{figure}[htbp]
\begin{center}
\begin{lstlisting}[style=MySCASP, basewidth=.42em]
claimStmt('C29', application(none), claim( application(none),[arducopter_software], 
                    [fit_for_purpose],[arducopterEnv] ), 'ArduCopter Software $\dots$') :-
  claimStmt('C30', application(none), claim( application(none),[arducopter_software], 
                    [posses_overarching_properties],[arducopterEnv] ), $\dots$),
  side_ClaimStmt('S30', application(none), claim( application(none),[overarching_properties], 
                    [explored_as_means_of_compliance],[arducopterEnv] ), 'The use of $\dots$'),
  claim(application(none), [arducopter_software], [fit_for_purpose], [arducopterEnv]).              
\end{lstlisting}
 
    \caption{Snippet of export of Arducopter Assurance Case into ASP.}
    \label{fig:PrologExport}
\end{center}
\vspace{-0.1in}
\end{figure}

When the ASCE tool exports the assurance case as ASP predicates, they are categorically saved in separate files: (i) the top-level claim is saved as a query, (ii) the negation of the top-level claim is saved as a negative query, (iii) the body of the assurance case are saved together as rules and facts, (iv) theory definitions are saved separately, and (v) defeaters are saved as integrity constraints that counter the claims. In the following section, we elaborate on various properties of interest and their analyses.

\subsection{Modeling Semantic Properties}
\label{sec:semant-prop-defin}


Semantic properties are rules about the contextual meaning of the concepts used in the statements within assurance cases such as consistency among claims, correctness of arguments, adequacy of evidence, etc. Although these properties are inherent in the minds of most authors and evaluators of assurance cases, to the best of our knowledge, they have not yet been systematically defined, let alone automatically checked. We outline some of the categories of properties that we have identified as crucial for evaluating the semantic rigor of assurance cases below. 

\begin{itemize}
\item \textit{Indefeasibly Justified}:  This property implies (a) the top-level claim is sufficiently supported by well-founded arguments and evidence, ensuring \textit{justification} and (b) there are no unresolved defeaters that could potentially alter the decision regarding the top-level claim called \textit{indefeasibility}. This semantic property is fundamental to an assurance case and in simple terms, it means the top-level claim can be indisputably deduced given the arguments and evidence.  
        
\item \textit{Theory Application Correctness}: Theories in Assurance 2.0 are reusable assurance fragments that can be independently specified, and `pre-certified' that could be applied to concrete assurance cases as sub-cases. However, when using theories, it is critical to ensure that they are instantiated appropriately and that all the necessary properties and evidences obligated by the theory are provided in the concrete case. This property guarantees the correct use of the theories. 
        
\item\textit{Property-Object-Environment Consistency}: Claims, arguments, and evidence assert properties of objects in specific environments. In extensive, hierarchically defined assurance cases, it is crucial to ensure no conflicts or contradictions exist among these assertions. While some conflicting terms are universally recognized, such as `X is safe' and `X is hazardous,' others, like `X has no vulnerabilities' and `X has residual security risks,' are domain-specific. We call these sets of inconsistent combinations of properties, objects, and environments as consistency rules for assurance.
        
\item \textit{Adequacy}:
  Assessing the sufficiency of sub-claims and
  evidence supporting the stated claims is a crucial aspect
  scrutinized in assurance cases. For example, for the Arducopter case we
  defined rules that required properties of evidence such as:
  \begin{itemize}
  \item[] \code{requirements_testcase_coverage_achieved},
  \item[]     \code{requirementsbased_testcases_passed} and
  \item[]     \code{structuralcoverage_of_requirementsbased_tests_achieved}
  \end{itemize}
  to be present to meet the claim with property:
  \begin{itemize}
  \item []\code{do178C_requirements_test_conformance_achieved}
  \end{itemize}
  Similarly, \code{meets_intent}, \code{is_correct}, and
  \code{innocuous} properties were required for the top-level claim
  with the \code{overarching} property. Any violation of these rules shows
  inadequacy in the assurance case.  Though these adequacy rules
  require a deeper understanding of the domain and context, unlike
  consistency rules, once defined, these adequacy rules will enable
  easy, rigorous, and recurrent validation of subsequent versions of
  large and complex assurance cases.
                      
      \item \textit{Completeness}: Completeness of assurance cases refers to the state of encompassing all the necessary elements in the domain of objects, properties, and environments for the system in consideration. While adequacy property is defined to find the presence of desired properties of a certain object, completeness concerns the presence of the same property for all the objects of a certain type. For instance, in the Arducopter case example, we defined rules that required the process of assessment to be completed (\code{process complete}) for all types of \code{assessments} performed on the Arducopter system. 
        
     \item \textit{Harmonious Coexistence of Theories}: As outlined in Section~\ref{sec:background}, Assurance 2.0 permits the incorporation of theories into an assurance case. However, employing multiple theories concurrently poses a risk of conflicts and contradictions stemming from disparities in their definitions or application methods, despite each theory being flawless on its own. Since these conflicts are mainly semantic, defining incompatible combinations as rules allows automatically checking for their presence. 
    \end{itemize}

The properties listed above represent only a fraction of potential assurance case properties. We view this effort as an initial step in identifying essential property categories for evaluating assurance case rigor. 
In the following section, we formally define these properties and discuss their automated analysis.



\subsection{Semantic Reasoning using s(CASP)}
\label{sec:semant-analys-using}

\begin{figure}[h]
\begin{center}
    \includegraphics[width=0.85\columnwidth]{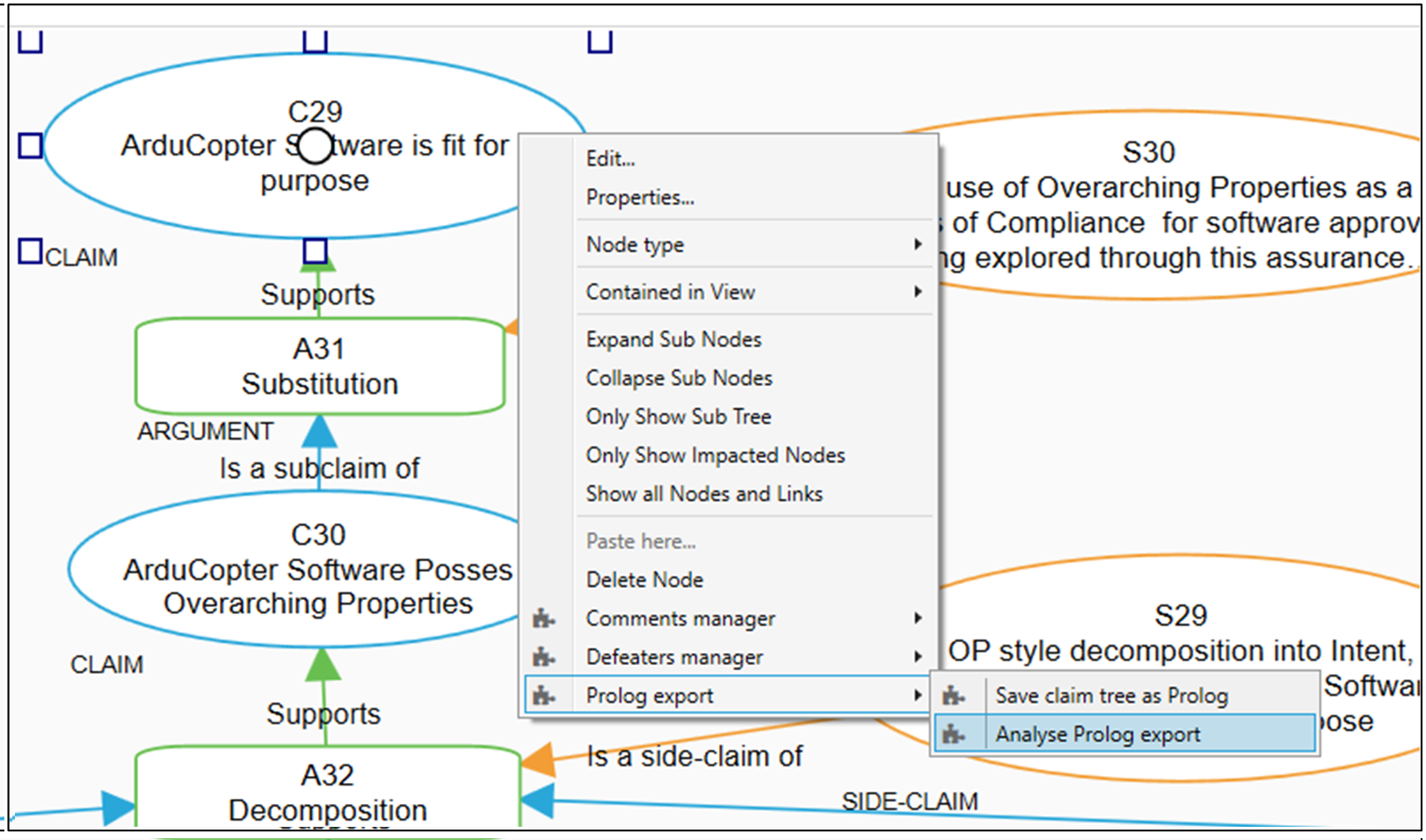}
    \caption{Semantic Analysis Option in ASCE Interface}
    \label{fig:ASCESemAnalysis}
    \end{center}
\end{figure}

In this section, we outline how we use s(CASP) to automatically analyze these properties using the exported ASP program of the assurance case. The ASCE tool is enhanced with semantic analysis capabilities, allowing users to invoke s(CASP) and perform semantic analysis through the ASCE interface, as illustrated in Figure~\ref{fig:ASCESemAnalysis}.

In essence, the s(CASP) system rigorously analyzes whether a specific query can be deduced in the provided ASP program. A successful execution yields the display of a ``model'', offering a detailed explanation as to why the query is entailed. Conversely, if the query cannot be deduced, executing the negation of that same query enables retrieving an explanation for the cause of the failure. Our objective is to harness this capability of s(CASP) to analyze the exported program of the assurance case for various semantic rules. Furthermore, for enhanced usability, we utilized the \code{--html} flag option of s(CASP) that displays the justification tree via an interactive HTML page, allowing the display of analysis results and models in a web browser. To address scalability concerns with the s(CASP) system, we enhanced the s(CASP) system l by implementing a more efficient and robust search, adding a debugger, and incorporating several builtins. 

In the remainder of the section, we delve into the details of analyzing each of the previously outlined properties using s(CASP), illustrated with examples from the Arducopter system. Although the specific properties and results are particular to this case example, the general methodology is widely applicable to a broad range of assurance cases.

\begin{figure}[h]
\begin{center}
    \includegraphics[width=1\columnwidth]{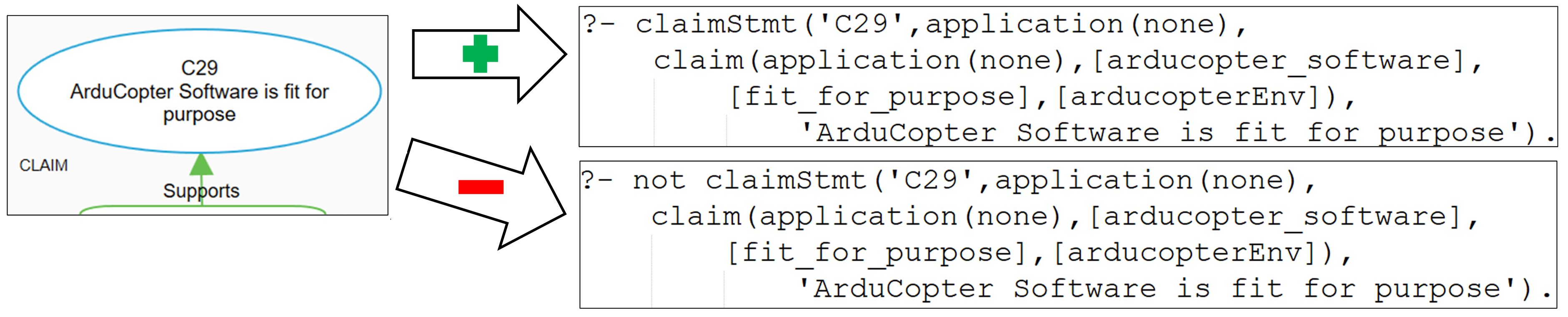}
    \caption{Positive and Negative Query Exported from Top-level Claim Node}
    \label{fig:query}
    \end{center}
\end{figure}

\paragraph{\bf Indefeasibly Justified:}
As mentioned earlier, this property concerns indisputable entailment of top-level claim. Formally, we represent this property as a query that includes the top-level claim. We have enhanced the ASCE tool to automatically and separately export the top-level claim in ASP as both positive and negative queries in addition to exporting the entire assurance case along with defeaters. The snippet of the exported query of ArduCopter assurance case’s top-level claim is shown in Figure~\ref{fig:query}.

\begin{figure}[h]
\begin{center}
    \includegraphics[width=\columnwidth]{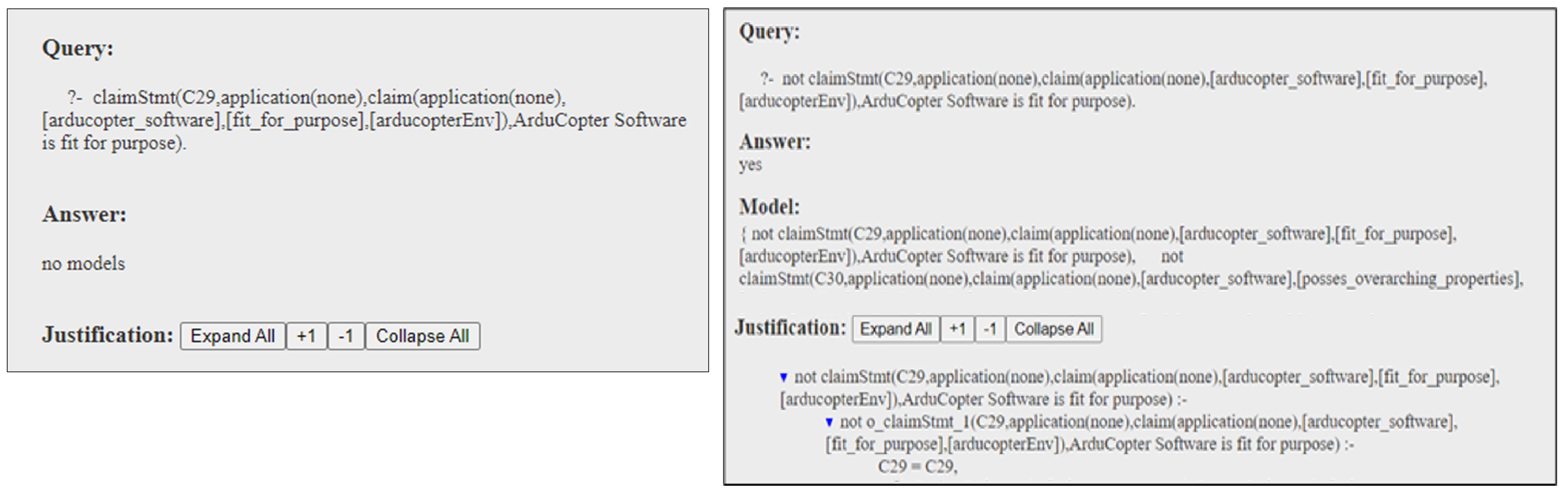}
    \caption{Example of s(CASP) Output of Failed Positive Query and Successful Negative Query}
    \label{fig:IDexec}
\end{center} 
\end{figure}

 When the positive query is successfully executed in s(CASP) and an explanation is provided, it signifies that the assurance case indeed possesses this property. Conversely, if the negative query is successfully executed, s(CASP) returns the unresolved defeaters as violations, as illustrated in Figure~\ref{fig:IDexec}. While the specifics of the model and justification depend on the system being analyzed, Figure~\ref{fig:IDexec} primarily shows how the results of positive and negative queries will be displayed.

\paragraph{\bf Theory Application Correctness:}
As previously explained, a theory is a reusable assurance case fragment applicable to concrete system assurance cases. To ensure reusability, objects and environments in theory nodes are defined as  variables (uppercase), while properties are  atoms (lowercase). Applying a theory in the concrete case requires the properties in that node and its sub-nodes to match the respective theory node's property. Additionally, the objects and environment of those nodes must be  atoms defined as instances of the respective  variables in the theory node. When authoring the assurance case, the types of objects and environments, along with their system-specific instantiations, should be predefined by the author. Using the ASCE interface, the author selects the desired theory and the correct instantiations (from the predefined list of type-instances) for each of the theory.

To automate the verification of correct theory application, we rely on s(CASP) to assess : (a) the direct match between properties outlined in the theory and those in the corresponding application nodes, and (b) the validity of all objects and environments as instantiations of the types specified in the theory nodes. For the analysis, we execute using the \textit{scasp} command along with the  exported program of the assurance case, the theory definition, and the same queries used to check the indefeasibly justified property. The analysis results are reported similarly to indefeasibly justified property analysis. 

\begin{figure}[h]
\begin{center}
    \includegraphics[width=\columnwidth]{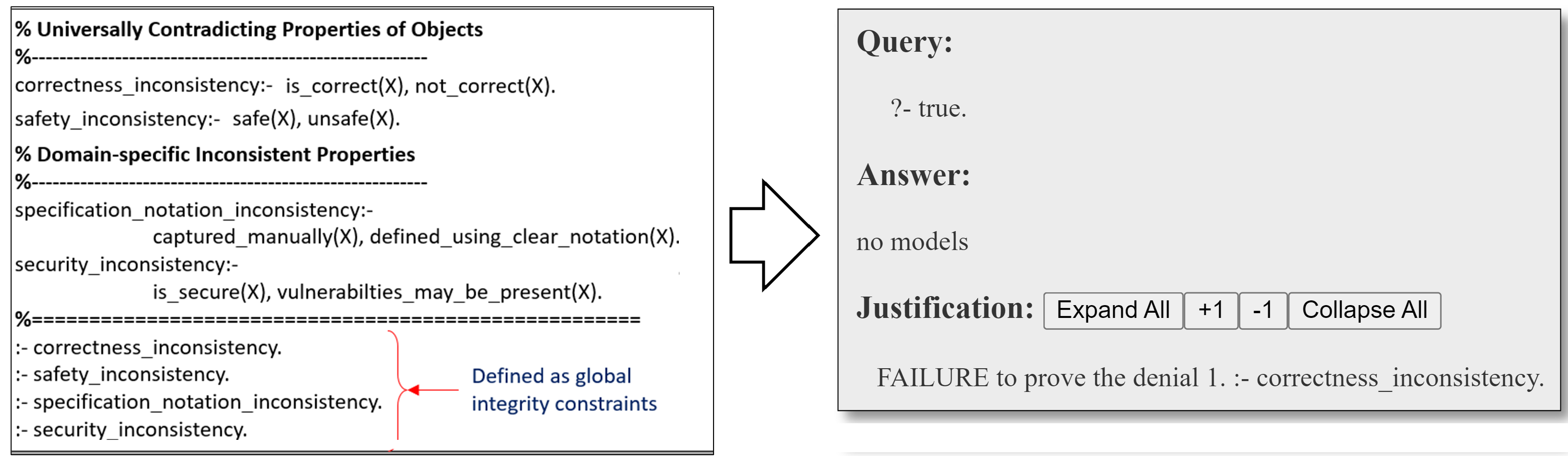}
    \caption{Consistency Rules and s(CASP) Analysis Output}
    \label{fig:Consistency}
    \end{center}
    \vspace{-0.15in}
\end{figure}

\paragraph{\bf Property-Object-Environment Consistency:}
We formally expressed consistency rules as global integrity constraints expressed in ASP notation. These constraints are specified in the form of conjunctions of \code{properties(Objects, Environment)} or \code{properties(Objects)}, where properties are  atoms, whereas Objects and environment are defined as  variables. This approach allows us to detect any consistency issues present in any instance of object or environment. The snippet displayed on the right side of Figure~\ref{fig:Consistency} illustrates the consistency rules devised for examining the ArduCopter case. 

To verify the adherence to these rules, we execute the \textit{scasp} command alongside the  exported program of the assurance case, the consistency rules, and the query, which is `?- true'. Essentially, this query prompts the s(CASP) engine to determine whether the assurance case contains instantiations (objects) for the inconsistent set of properties defined by the rules. s(CASP) notifies us of any violations it discovers, as demonstrated on the left side of Figure~\ref{fig:Consistency} for the Arducopter case. 

\paragraph{\bf Adequacy:}
The adequacy property is also specified as  rules structured as conjunctions of  \code{properties(Objects, Environment)} or \code{properties(Objects)}, similar to consistency rules. While consistency rules are global integrity constraints, adequacy rules are designed to verify the concurrent presence of properties for the same instance of an object. Therefore, the query posed to s(CASP) is whether all the properties in the conjunction are present in the assurance case. We execute the \textit{scasp} command using the  exported program of the assurance case, the adequacy rules, and this query. Essentially, this query prompts the s(CASP) engine to verify whether the assurance case contains instantiations (objects) with all the properties.

\begin{figure}[h]
\begin{center}
    \includegraphics[width=1\columnwidth]{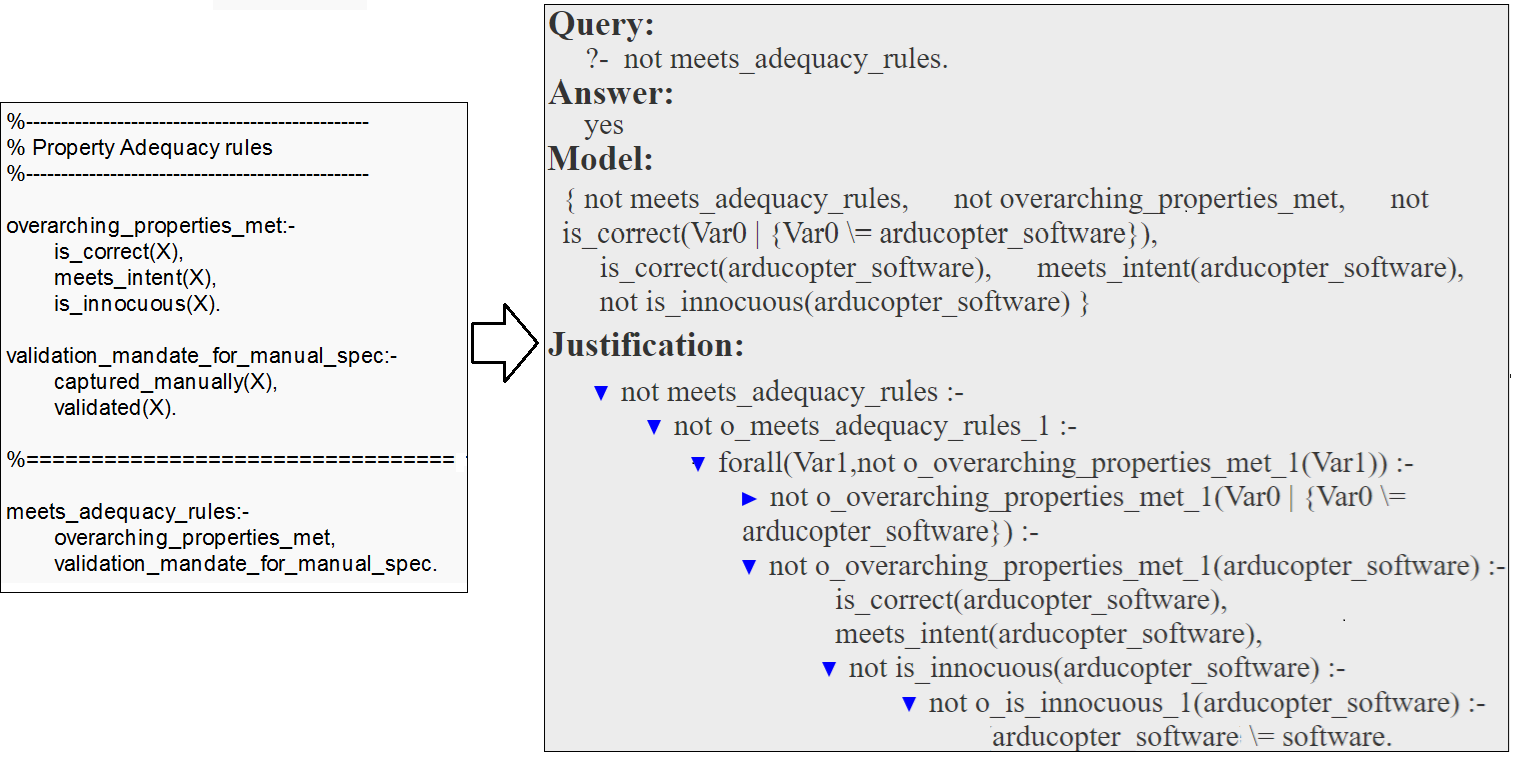}
    \caption{Adequacy Rules and s(CASP) Analysis Output}
    \label{fig:Adequacy}
    \end{center}
\end{figure}


For instance, consider one of the adequacy rules defined for ArduCopter, which checks if the three overarching properties, namely \code{meets_intent}, \code{is_correct}, and \code{innocuous}, are satisfied, as shown in Figure~\ref{fig:Adequacy}. When analyzed using the  exported program of the Arducopter case, the s(CASP) engine searches through the case to identify instantiations that fulfill this rule. If this condition is not met, as illustrated on the right side of Figure~\ref{fig:Adequacy} for explanatory purposes, by negating the query to s(CASP), we get a justification tree, detailing the reason for the `inadequacy', such as the absence of the \code{is_innocuous} property for the \code{arducopter_software} object.

\paragraph{\bf Completeness:}
Completeness properties concern the domain of the objects, properties and environment within the entirety of the assurance case. Since Assurance 2.0 case creation requires authors to define a global set of objects, properties, and environments for the system under consideration, we utilize this predefined set for specifying and analyzing completeness properties. These completeness property specifications are similar to adequacy. However, instead of verifying if completeness is met, we negate the property and query s(CASP) to determine the reason why the assurance case does not meet the completeness property. This query prompts s(CASP) to search for an instantiation of an object that does not satisfy completeness. This level of detail allows the assurance evaluator to assess if all instantiations of a certain type have certain common properties.

\begin{figure}[h]
\begin{center}
    \includegraphics[width=1\columnwidth]{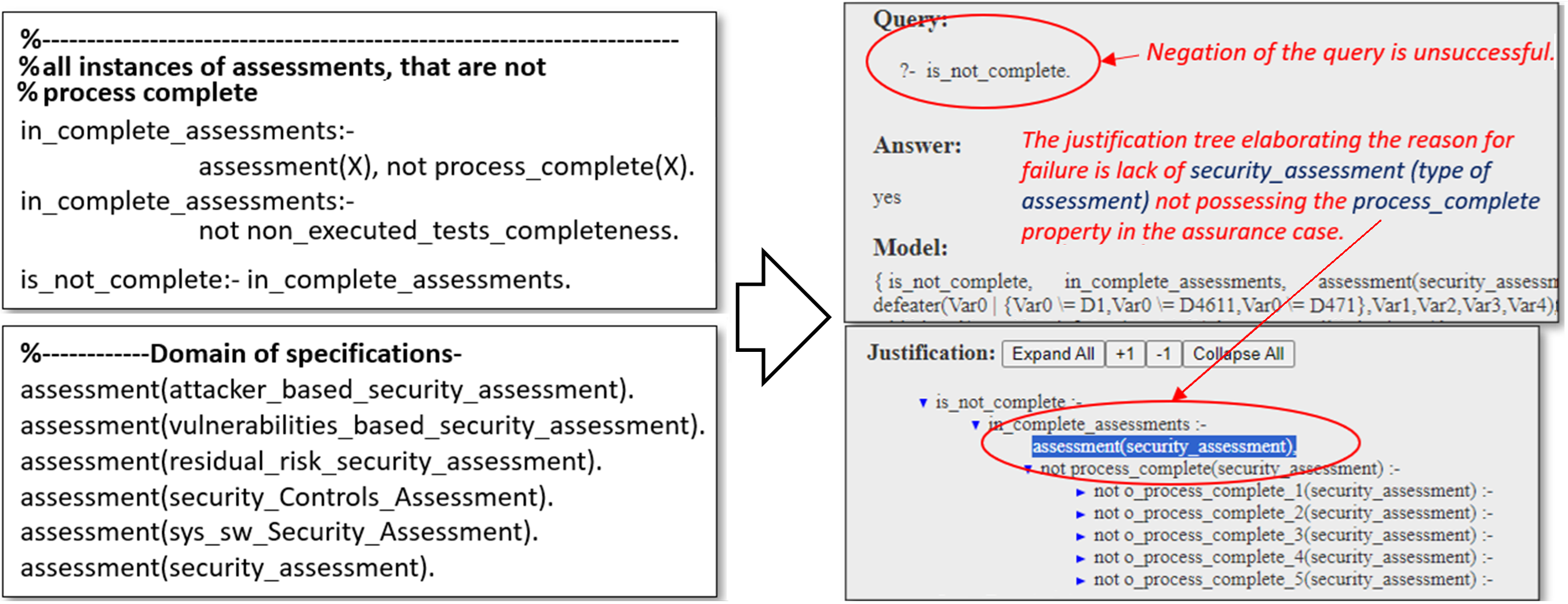}
    \caption{Completeness Rules and s(CASP) Analysis Outputs}
    \label{fig:Completeness}
    \end{center}
    \vspace{-0.15in}
\end{figure}

In the Arducopter case, we established a completeness property \code{assessment(X)} and \code{not process_complete(X)}. When the negation of this rule was executed in s(CASP), along with the assurance case and the definition of the domain types and their instantiations, the result was a justification explaining the reason for the failure. As illustrated in the figure, the lack of the \code{process_complete} property in the assurance case for the \code{security_assessment} instance of type \code{assessment} was identified as the cause of failure.

\paragraph{\bf Harmonious Coexistence of Theories:} 
Inharmonious theory definitions of rules resemble consistency rule definitions. For example, in the Arducopter case, we defined a rule for \code{inharmonious_DAL_theories}, to identify if theories  pertaining different DAL levels coexist, such as \code{achieves_DAL_C_DO178c_requirement_testing(X)} and \code{achieves_DAL_A_DO178c_code_coverage(X)}. These are also defined as global integrity constraints, and their analysis is conducted in the same manner as consistency rules. Essentially, this query prompts the s(CASP) engine to determine whether the assurance case contains references to theories that are defined as inharmonious, as defined by the rules. s(CASP) notifies us of any violations it discovers, similar to the way the inconsistencies were reported (as shown earlier in Figure~\ref{fig:Consistency}).

In sum, the realm of semantic analysis offers a wide range of possible analyses. As part of our ongoing work, we are exploring valuable and powerful properties to analyze. Automating these analyses provides valuable insights and relieves humans from repetitive tasks, leading to improved decision-making regarding assurance cases.


\vspace{-0.1in}

\section{Conclusion}
The Assurance 2.0 framework aims to advance the science behind assurance cases and enhance confidence in their development and assessment across various certification regimes. This paper introduces our method of enriching the Assurance 2.0 framework with semantic analysis capabilities by harnessing the reasoning abilities of s(CASP). Our approach innovatively involves systematically translating Assurance 2.0 cases into Answer Set Programming (ASP) notation and formally defining key properties essential for the robustness of assurance cases, thereby enabling semantic analysis via s(CASP). To our knowledge, this methodology represents the first effort to automate the semantic analysis of system assurance cases, establishing a novel paradigm for explainable, knowledge-assisted reasoning. Evaluations conducted on industrial-strength case studies in safety-critical domains such as avionics and nuclear reactors have yielded positive feedback from assurance authors and evaluators. As part of future work, we intend to explore the use of LLMs for semantic property specification and analysis, as well as enhance the tool's capability to analyze large assurance cases assembled using several complex theories. While this paper focuses on the semantic analysis of assurance cases within our CLARISSA approach, interested readers are referred to \cite{finalCLARISSAReport} and \cite{SASSUR2024} for a comprehensive overview of our work.

\section*{Acknowledgements}
CLARISSA is supported by DARPA contract number FA875020C0512. Distribution Statement ``A'': Approved for Public Release, Distribution Unlimited. The views, opinions, and/or findings expressed are those of the authors and should not be interpreted as representing the official views or policies of the Dept of Defense or the U.S. Govt. 

We are grateful to the anonymous reviewers for their insightful comments and suggestions for improvement.

\bibliographystyle{tlplike}
\bibliography{clarissa}

\end{document}